# Multi-level Trust based Intelligence Schema for Securing of Internet of Things (IoT) Against Security Threats Using Cryptographic Authentication

Kobra Mabodi[1] . Mehdi Yusefi [2] . Shahram Zandiyan[3] . Leili Irankhah[4] . Reza Fotohi[5]

**Abstract** The Internet of Things (IoT) is able to provide a prediction of linked, universal, and smart nodes that have autonomous interaction when they present services. Because of wide openness, relatively high processing power, and wide distribution of IoT things, they are ideal for attacks of the gray hole. In the gray hole attack, the attacker fakes itself as the shortest path to the destination that is a thing here. This causes the routing packets don't reach the destination. The proposed method is based on the AODV routing protocol and is presented under the MTISS-IoT name which means for the reduction of gray hole attacks using check node information. In this paper, a hybrid approach is proposed based on cryptographic authentication. The proposed approach consists of four Phases, such as the verifying node trust in the IoT, testing the routes, gray hole attack discovery, and the malicious attack elimination process in MTISS-IoT. The method is evaluated here via extensive simulations carried out in the NS-3 environment. The experimental results of four scenarios demonstrated that the MTISS-IoT method can achieve a false-positive rate of 14.104 percent, a false-negative rate of 17.49 percent and a detection rate of 94.5 percent when gray hole attack was launched.

**Keywords** Internet of Things (IoT) . Gray hole attack . Multi-level trust . Cryptographic authentication

✉ Kobra Mabodi
   Maboudi.79@gmail.com

✉ Mehdi Yusefi
   Mehdi93757@gmail.com

✉ Shahram Zandiyan
   Sh.zandian@iauardabil.ac.ir

✉ Leili Irankhah
   L_irankhah@yahoo.com

✉ Reza Fotohi
   R_fotohi@sbu.ac.ir;  Fotohi.reza@gmail.com

[1] Department of Computer Engineering, Non-profit higher education institutions, Ahvaz, Iran.
[2] Department of Computer Engineering, Islamic Azad University, Dolatabad Branch, Isfahan, Iran.
[3] Department of Computer. Ardabil Science and Research Branch. Islamic Azad University. Ardabil. Iran
[4] Department of IT and Computer Engineering, Urmia University of Technology, Urmia, Iran
[5] Faculty of Computer Science and Engineering, Shahid Beheshti University, Tehran, Iran

# 1 Introduction

IoT forms a system through interconnecting various machines, devices, and software services. IoT can increase the quality of human life and play a significant role in today's modern life since through energy-efficient automation. It should be noted that since IoT systems are resource-constrained and have an ad-hoc nature can be influenced by cyber attackers [1]. This necessitates designing reliable and secure IoT, and the challenges should be overcome so that other systems and human lives are not damaged or destroyed. Some attacks like a gray hole attack have illegal penetration into the system. When an attack affects an IoT, it is difficult to remove the threat and bring the system back online. It should be noted that the conventional approaches for securing information like intrusion detection or encryption are not adequate when these risks are dealt. The sensor and actuator measurements compatibility factor with IoT control mechanism and the physical process are not taken into account in these plans. These cases are essential for the protection scheme. Due to the high volume of sensing and network data generated by IoT systems and devices, it is very effective to have cryptographic authentication in constant monitoring and analysis for the IoT systems' security. For instance, as observed in Fig. 1, a gray hole attacker can strive to imitate the wireless router connecting the IoT devices with the rest of the network. Then, it can reroute the traffic by the legal wireless router.

**Fig. 1** IoT gray hole attacks based on application scenarios

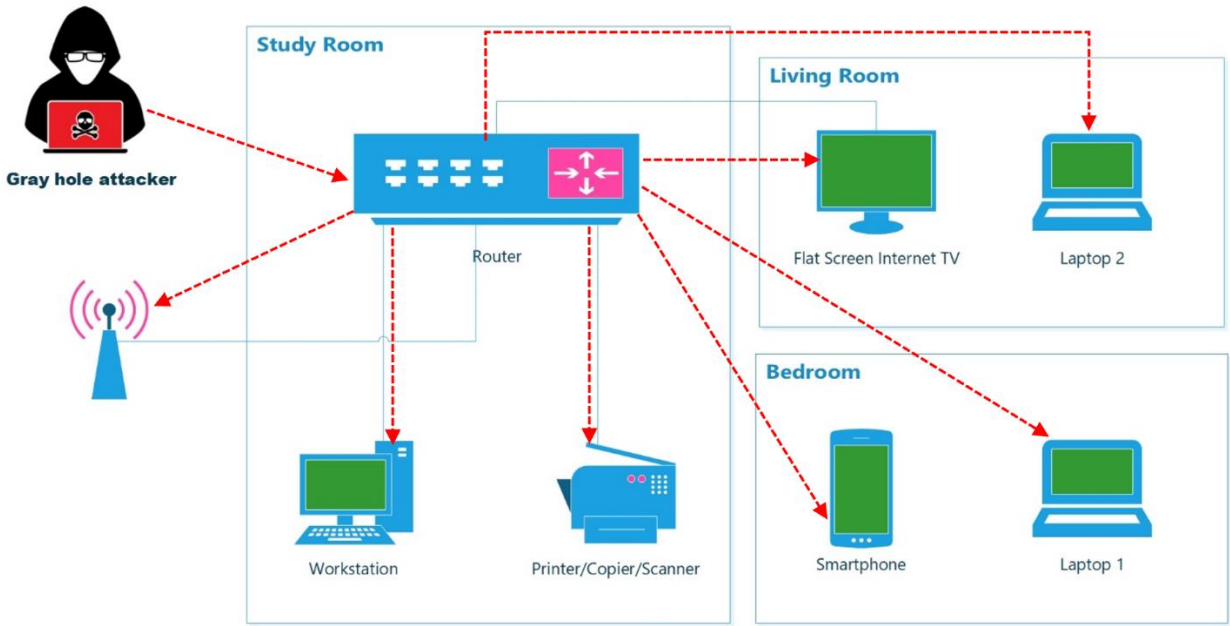

In this paper, a gray hole attack discovery and prevention method via informing other nodes in the internet of things is proposed. The proposed scheme first verifies the trust level for the nodes in the network and then discovers and eliminates the malicious gray hole nodes using control packets. The proposed method is based on the AODV routing protocol and is presented under the MTISS-IoT name which stands for the reduction of gray hole attacks using check node information.

The paper presented here is organized as the following. Section 2 converses the security attacks for IoT networks. The related work is discussed in Section 3. In Sect. 4 brings the proposed MTISS-IoT strategy. In Section 5, the simulation results are discussed to demonstrate the efficiency of the proposed MTISS-IoT. Finally, conclusions and future works of this research are discussed in Section 6.

## 2 Security attacks

IoT Systems are vulnerable to function degradation and security risks. They might be passive or active threats as they have the reliance on wireless channels for communication. A security threat targeting IoT is provided in Figures 2 and 3. In this paper, the following vulnerability is of interest:

***Gray hole Attack***: In a gray hole attack, there is a malicious node among the selected path nodes that drops the messages and don't forward them to the adjacent nodes. In this regard, two possibilities can be mentioned. In the first possibility, the gray hole node fakes itself as a routing node by observing the routing protocol accurately. In another possibility, the attackers don't observe the routing protocol and so that it violates the protocol specifications by exploiting its vulnerabilities. The gray hole node fakes itself as a routing node in different ways based on the used protocol. The gray hole attacks can be divided into two groups based on the number of malicious nodes—simple and cooperative gray hole [2].

***Simple gray hole attack:*** In this kind of gray hole attack, a malicious node foists itself as a medium node that belongs to the shortest route to the destination. Regardless of the routing table, the gray hole node is always accessible to reply route requests to receive data packets and drop them instead of forwarding them. In the flooding-based protocols, before sending a reply by healthy nodes, the gray hole node sends a reply to the requesting node. In this way, the selected route will contain a malicious node which drops the packets or send them to incorrect nodes. The process of a gray hole attack is shown in Fig. 2. As seen in this figure, the data packets should be transferred from a source $Th_S$ to a destination $Th_D$. For this purpose, a proper route—from the origin to the destination—should be detected. So, if $Th_M$ is malicious, it fakes itself as a node present in the shortest path to the destination. Then, it will respond to the request by sending a reply to the $Th_S$ sooner than other nodes. in this way, the $Th_S$ will send the data packets to $Th_M$ and discard the replies received from other nodes [3].

**Fig. 2** Simple gray hole attack.

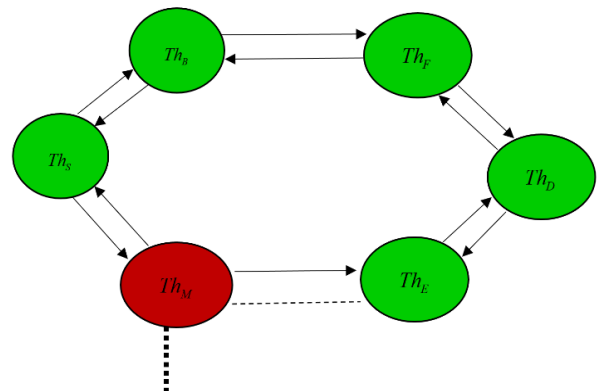

***Cooperative gray hole attack: C***ooperative gray hole attack is the other way of implementing a gray hole attack where there is cooperation among different malicious nodes in the attack for violating the security system and routing protocol. As indicated in Fig. 3, when malicious $Th_{M1}$ and $Th_{M2}$ cooperate, $Th_{M1}$ refers to $Th_{M2}$ as its subsequent hop. Given the scenario described in [4], the source $Thing_S$ transfers a Further Request packet ($FReq_{packet}$) to $Th_{M2}$ by another route other than via $Th_{M1}$ (for example, $Th_S - Th_C - Th_E - Th_{M2}$). The $Thing_S$ requests $Th_{M2}$ if it is the subsequent hop of $Th_{M1}$ and if it has a reliable path to the destination $Th_D$. As there is a cooperation between the $Th_{M2}$ and $Th_{M1}$, its Further Reply ($FReq_{packet}$) is positive. Thus, it is assumed by the source $Th_S$ that the path $Th_S - Th_{M1} - Th_{M2}$ has security, and it begins transferring the data packets over it. $Th_{M1}$ will drop the packets following being intercepted.

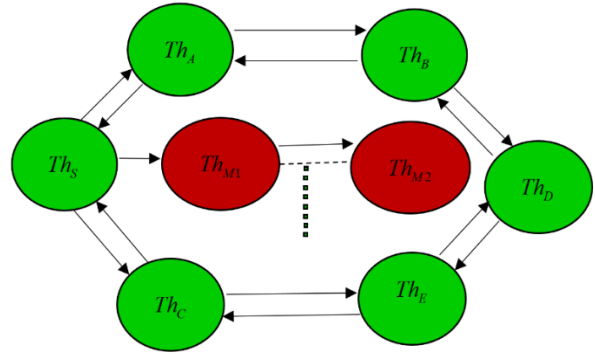

**Fig. 3** Cooperative gray hole attack.

## 3 Related works

Various security measurements have been developed and used in different works for addressing Denial of Sleep attacks, and protecting IoTs against gray hole attacks. It is not a recent issue, and there are extensive studies on it. Different approaches have been suggested by different studies in order to address these attacks.

In this paper, a new real-time hybrid intrusion detecting frame-work is provided including specification-oriented and anomaly-based intrusion detecting modules to detect two well-known routing attacks in IoT known as selective-forwarding and sinkhole attacks. Hence, the host nodes' behavior is analyzed by the specification-oriented intrusion detecting elements situated in the router nodes, and their local outcomes are sent to the root node via normal data packets. Furthermore, the unverified optimal-path forest algorithm is used by an anomaly-oriented intrusion detection agent situated in the root node to project clustering models via incoming data packets. This agent that is on the basis of the MapReduce architecture is able to act in a distributed platform to project clustering models and therefore detect parallelly the anomalies as a global detection method. By the suggested technique, decisions are made regarding suspicious performance via a voting mechanism. Remarkably, the suggested technique is also prolonged for detecting wormhole attack. Deploying the hybrid suggested model is also assessed in a smart-city scenario via a present platform. The scale of the free network and the capability in identifying malevolent nodes are two key properties of the suggested framework that are assessed via various tests in this work [5].

Here, the authors provide an algorithm in terms of the exponential smoothing concept for detecting the nodes' topological isolation owing to blackhole attack. Exponential smoothing is a method to smooth time series data utilizing the exponential window function and it is utilized for short, medium and long-term prediction. We use the exponential smoothing in our suggested algorithm to approximate the next onset time of packets at the sink node from each of other nodes in the LLN. This estimation is used to design the algorithm for identifying the malevolent nodes initiating real blackhole attack [6].

In [7], SCOTRES—a trust-oriented system is proposed for secure routing in ad-hoc networks to advance the network entities' intelligence using 5 innovative metrics. The resource consumption of each node is considered by the energy metric to impose similar quantity of collaboration and to increase the network's lifetime. The topology metric knows the positions of the nodes and improves the load balancing. The tolerance in periodic malfunctioning is provided by channel-health metric owing to bad channel circumstances and the network is protected versus jamming attacks. The collaboration of each subject for a particular network operation is evaluated by reputation metric to detect the specific attacks, however, the total compliance is estimated by trust metric, protecting against combinatorial attacks. The system's security features are validated by the Theoretic analysis.

Blackhole and selective forwarding routing attacks are addressed in [8], which are the basis security attacks on the data routing in IoT networks. Today, most IoT tools, from medical instruments to connected vehicles and even smart buildings are able to communicate with each other wirelessly. In this work, a trust-oriented routing Protocol is provided for Lossy and Low-Power Networks stating blackhole and selective forwarding attacks. We indicate that our suggested protocol is secured from blackhole and selective forwarding attacks, and it does not enforce undue overheads on network traffic.

In [9], a comprehensive study of RPL, its known attacks are represented and the mitigation approaches are suggested to counter these attacks. We performed a complete review of the RPL standard, containing a currently suggested modification. Moreover, all recently published attacks on RPL and their mitigation approaches were investigated within the literature. According to this assessment, and as we know, this is a first-of-its-kind categorizing outline for the mitigation approaches in terms of the methods utilized for the mitigation. Moreover, we systematically deliberated RP-oriented Intrusion Detection Systems (IDSs) and their categorizations, while remarking the most lately suggested IDSs.

This work suggested constructing a trust-oriented framework for RPL to counter blackhole attacks. It can be run at two levels of an intra-DODAG and an inter-DODAG. Incremented dropping the packets, depleting the resources, and high packet overhead are the impacts of blackhole attacks in an IoT network. It eventually leads to destabilizing the network owing to incremented packet delay, rank modifying and disturbance in the topology. Regarding the rank modifying, the ranks are computed again, therefore, activating a local repair later initiating a repair thoroughly by the root. Such regular repairs might end up influencing the network efficiency [10].

In [11], SIEWE (Strainer based Intrusion Detection of Blackhole in 6LoWPAN for the Internet of Things) an Intrusion detection mechanism is proposed to recognize Blackhole attack on Routing protocol RPL in IoT. To arrange the Blackhole attack, a malevolent node first should broadcast a comparatively great routing metric to the nearby nodes so that it seems to be the best candidate chosen as a parent. The above fact is used by SIEWE to filter out the nodes broadcasting a comparatively great routing metric and appending their node IDs to a suspect list. Furthermore, detecting and verifying are performed by only the nodes with at least one entrance in their suspect list set. The above nodes analyze the nodes' behavior in their respective suspect lists while sending their observations to BR node. Therefore, SIEWE includes only those nodes in the vicinity of suspected nodes rather than involving every resource inhibited node in the network for detecting and verifying procedure. Hence, overall consumed energy is saved by SIEWE within the network and the number of observation packets moved to BR node is limited.

SPLIT, a secure and scalable RPL routing protocol is proposed in [12] for IoT networks. A lightweight remote attestation method is effectively utilized by SPLIT to guarantee the network nodes' software integrity. SPLIT piggybacks attestation are processed on the RPL's control messages to prevent further overhead resultant by attestation messages. Consequently, SPLIT benefits the RPL protocol's scalability propertied and low energy consumption that are essential for the resource-constrained large-scale networks like IoT. The simulation outcomes for various IoT setups indicate the

SPLIT effectiveness in comparison with the state-of-the-art by existing various kinds of attacks regarding metrics like energy consumption and packet delivery ratio.

One of the methods offered to detect and dispel a DoS attack in contrast to the IoT middleware—which is also known as NPS—is the REATO method. A real test-bed is used to authenticate the premeditated solution for NPS architecture. This solution is composed of an NPS sample mounted on a Raspberry Pi which receives open data feeds in real-time using an adaptable source set. To find a solution to detect DoS attacks in the IoT, it should be noted that we should consider all the potential circumstances—attacks to the data sources and IoT platform) [13].

A deep-learning established machine learning method has been presented in [14] for the IoT to detect the routing attacks. The Cooja IoT emulator has been employed to generate high-fidelity attack data within IoT networks having 10 to 1000 nodes. They have recommended a highly scalable, profound-learning based attack detection approach to uncover the IoT routing attacks which are decreased rank, hello-flood, and version number modification attacks through extraordinary accurateness and meticulousness. Applying the deep learning for cyber-security in the IoT necessitates the accessibility of considerable IoT attack data.

Qin et al. [15] proposed an IMLADS to manage the security of the IoT in a well-organized manner. This method differs from the traditional systems so that it employs the movable agents—instead of stationary one—to complete the data collection and analysis stages. in this way, the movable agent running platform controls the mobility despite the fact that that is irrelevant to its setup system. The data can be transferred to the other nodes based on a pre-set monitoring task. By taking advantage of this technology, the number of agents running in the system can be increased significantly while enhancing the steadiness and scalability of the system. Therefore, various approaches have been designed by the authors of the mentioned study for node and system-level security monitoring. In the first approach, they employed a lightweight data collection and analysis method. It should be noted that only small local computing resources can be dominated by this method. In the second approach, they proposed a parameter calculation technique. However, this approach has perpetual computational complexities.

The previous works to design IDS for the IoT have been listed in Table 1 ("-" indicates the indefinite characteristics).

**Table 1** Summary of the approaches for IoT literature.

| References | Detection schema | Attack type | Validation schema |
|---|---|---|---|
| [5] | Hybrid | Routing attacks | Simulation |
| [6] | Exponential smoothing | Blackhole Attacks | Simulation |
| [7] | Trust-based system | Blackhole &jamming attacks | Simulation |
| [8] | Trust-based routing | Blackhole & Selective-Forwarding Attacks | Simulation |
| [9] | Hybrid | Blackhole & Selective-Forward Attacks | none |
| [10] | Trust-based mechanism | Blackhole Attacks | Simulation |
| [11] | Anomaly-based | Blackhole Attacks | Simulation |
| [12] | Anomaly-based | Hybrid | Simulation |
| [13] | Hybrid | DoS attacks (blackhole, …) | Test-bed |
| [14] | deep-learning-based | cyber security attacks (blackhole, …) | Emprical |
| [15] | NOS-based | Routing attacks | Simulation |

# 4 The proposed MTISS-IoT approach

In the following section, we design a gray hole-security threats-immune schema by employing the cryptographic authentication. The MTISS-IoT consists of four Phases, such as the verifying thing trust in the IoT is discussed in Sect, 3.1. Testing the routes is discussed in Sect, 3.2., gray hole thing discovery in MTISS-IoT is discussed in Sect. 3.3, and the malicious thing elimination process in MTISS-IoT is discussed in Sect. 3.4.

## 4.1 The assumptions applied in the proposed MTISS-IoT

The proposed method works under the following assumptions:
- All of the things are identical in their physical characteristics.
- If thing X is in the transmission range of thing Y, then thing Y is in the transmission range of thing X too.
- All of the things have been verified and can take part in the transmissions.
- Both the source thing and the destination thing are IDS things that are used to detect the malicious thing along the route.

## 4.2 Phase 1: Verifying thing trust in the IoT

In the proposed MTISS-IoT method, each node monitors the information regarding its immediate neighbors, the nodes within a single step distance, and also other nodes in the network using RREQ packets. In the proposed MTISS-IoT method, the route discovery process is the same as the base AODV method. If the source node needs to discover a route to reach the destination, it broadcasts the RREQ packet. According to our proposed method, once an intermediate node receives the RREQ packet for a specific destination, it carries out the following actions:

For each intermediate node on the RREQ packet route, once receiving this message from a node, it increases the ID of that node and its RREQ_C credit field by one (the initial RREQ_C credit value is equal to zero) and inserts it into its monitoring table. Then checks whether the RREQ packet is duplicate (checks to see if its own ID is listed in the source route). If this is the case, the node drops that RREQ packet. Otherwise, if the node has a route to the destination then it submits its request into the RREQ packet and sends an RREP packet back to the source node. However, if this is not the case and the node does not have a route to the destination, then the node adds its ID to the source route and broadcasts the RREQ packet in the network. After sending the RREQ packet to its neighboring nodes, the node transfer field (RREQ_T), which has an initial value of zero, is increased by one. The RREQ_C credit field for a specific node is increased by its adjacent nodes every time it forwards an RREQ packet. Each time a node transmits a packet to one of its neighboring nodes, the RREQ_T field for that neighbor is increased by one. This way, each node can collect more information about the behavior of other nodes during the route discovery process. Each node sets its trust to other nodes to either high or low according to the value of the RREQ_C credit field. The monitoring table is also run periodically to represent the current topology and the behavior of the network nodes. Table 2, 3, and 4 demonstrates an example of a monitoring table for Figure 4.

**Fig. 4** An example of thing layout in the proximity of a suspicious thing in an IoT.

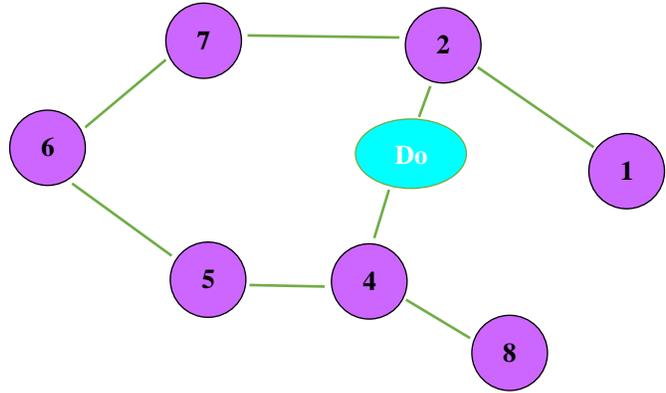

**Table 2** An example of a monitoring table for Thing 4.

| Thing 4 | RREQ_T | RREQ_C |
|---|---|---|
| 5 | 0 | 0 |
| 8 | 0 | 1 |
| D0 | 1 | 0 |

**Table 3** An example of a monitoring table for Thing 7.

| Thing 7 | RREQ_T | RREQ_C |
|---|---|---|
| 6 | 0 | 1 |
| 2 | 1 | 0 |
| D0 | 1 | 0 |

**Table 4** An example of a monitoring table for Thing Do.

| Thing Do | RREQ_T | RREQ_C |
|---|---|---|
| 7 | 0 | 1 |
| 4 | 0 | 1 |
| 2 | 0 | 0 |
| 5 | 1 | 0 |

As presented in these tables, the Do node has received two different RREQ packets from nodes 4 and 7 but has not sent any packets to node 2. In the meanwhile, it has only sent one packet to node 5 and eliminated some packets selectively. While other nodes have sent the packet to their neighbors normally after receiving it. For instance, node 4 has sent the packet to node Do after receiving it from node 8. The trust value X of node X for node Y is calculated using Eq. (1):

$$T(X,Y) = \text{The calculated RREQ\_C credit by thing A for thing B} \tag{1}$$

According to the AODV routing protocol, a node can eliminate a received *RREQ* packet only in two cases. First, when it receives an *RREQ* packet that has been received before and the node has already broadcast it. Second, when the node receives an *RREQ* which contains a new route to the requested destination. In this case, the node deletes the *RREQ* packet and sends an RREP to the source node. The trust threshold value for a neighbor is represented by $T_c$. Selecting the optimal threshold value is highly valuable for ensuring the good performance of the network. We analyze this value by testing the *RREQ* packets and with regard to the behavior of the nodes in the network. Node *X* is trustworthy for node *Y* if the *RREQ_C* and *RREQ_T* columns for node X in the monitoring table of node *Y* are identical. In other words, it transmits every unique *RREQ* it receives to its immediate neighbor.

For our analysis, we considered a threshold $T_c$ which is the difference between the RREQ_C and *RREQ_T* values for a specific node and its neighbors. Since the gray hole nodes eliminate some forwarded *RREQ* packets, we set the threshold value equal to zero for the nodes that are considered normal. This is because the nodes are supposed to transmit every unique *RREQ* packet they receive. In fact, the threshold value is calculated by subtracting the number of *RREQ* packets transmitted from a specific node from the number of *RREQ* packets transmitted to that node (*RREQ_T*). Eq. (2) demonstrates the $T_C$ calculation.

$$T_C = \left(\sum RREQ\_T - \sum RREQ\_C\right) \tag{2}$$

Just as the monitoring table is updated periodically, the behavior of each node is also observed by other nodes at specific time intervals. Each node needs to transmit the last *RREQ* packet it has received at different time intervals. Therefore, we say that the trust value for a node is high if the trust field of that node is equal to the threshold value, in other words, it sends all of the packets. On the other hand, we say that the trust value for a node is low if the trust field for that node is higher than the threshold value, i.e., it does not send some of the packets. Eq. (3) and Eq. (4) demonstrates this calculation.

*The trust value of* thing *X for* thing *Y is high if* $: T(X,Y) = T_C$ (3)

*The trust value of* thing *X for* thing *Y is high if* $: T(X,Y) > T_C$ (4)

In Figure 4, the neighboring nodes of node Do[4] (the suspicious node) are nodes 2 and 4. We assume that nodes 2 and 4 are normal nodes and node Do is suspected to be a gray hole node. Every node except Do will take part in the *RREQ* packet transmission process normally. However, node Do will delete some of the transmitted *RREQ* packets and therefore its trust value will be higher than zero ($T_C$). The trust value for the intermediate nodes and the neighboring nodes of the suspicious node are presented in Table 5 These values were calculated after sending the monitoring table and with the trust threshold equal to zero. The behavior of the nodes after receiving an RREQ packet is shown in Fig 5.

**Table 5** Trust table.

| Node | Calculated trust value based on the RREQ_T and RREQ_C fields |
|---|---|
| 1 | High |
| 2 | High |
| Do | Low |
| 4 | High |
| 5 | High |

---

[4] Doubtful

**Fig. 5** The behavior of the nodes after receiving an RREQ packet

| **Neighboring node behavior monitoring algorithm when it receives an RREQ packet:** |
|---|
| 1. If the receiving node is the destination IDS node: |
|     i. Highest trust value for all of the nodes along the requested route is calculated |
|     ii. RREP is sent back to the source node through the route with the highest trust value |
| 2. Otherwise, if the node is not the destination: |
|     i. If the node ID is not in the source route of the RREQ packet, all of the node IDs in the RREQ route are inserted into its monitoring table. |
|     ii. The RREQ_C credit values are increased for all of the sender nodes and by sending the RREQ message forward, RREQ_T value for all of the receiving nodes is also increased |
| 3. Otherwise, if RREQ is duplicate: |
|     i. Delete the RREQ packet and insert the IDs of the neighboring nodes which have transmitted the duplicate RREQ packet |
|     ii. Increase the RREQ_C credit value of that node in the monitoring table |

### 4.3 Phase 2: Testing the routes

Once an intermediate node receives a *RREP* packet, it copies the trust value assigned to the next node on the route to the destination. Afterward, it sends that packet to the previous node and this process is repeated until the *RREP* packet reaches the source node. According to the algorithm presented in Figure 7, all of the current nodes on the route from the source to the destination which send the response packet, increase the trust value for their next node in the *RREP* packet. When the source node receives the first *RREP* packet, it checks the corresponding *RREQ* packets to see if the response is from the destination or from intermediate nodes. If the response is from the destination, then it considers the route for packet transmission. However, if the received response is from an intermediate node claiming to have a new route to the destination, then the source node selects the *RREP* packet which has the highest trust value assigned to the nodes along its corresponding route. Our method, on the other hand, adds one more checkpoint for discovering malicious nodes in addition to discovering highly trustworthy routes.

Once the source node receives the *RREP* packet, it sends a test packet comprised of multiple blocks to the destination along the routes with the highest trust values. The destination is bound to send an acknowledgment packet for that route which includes the number of the received blocks. Each intermediate node in the route needs to count the number of blocks in the test packet it hands over to the next node in a Number of Packets Forward manner. When the destination node receives the data packets from the source node, it starts counting and stores the number of packets in the received data in a block. Obviously, if a route is infected by a gray hole node then the test packet will either not reach the destination at all or it will not reach the destination in its entirety. The destination IDS node sends the number of blocks in the test packet to the source node using the acknowledgment packet. Once the source IDS node receives the acknowledgment packet, it checks to see if the number of blocks is the same as the number of blocks in the test packet. If fewer blocks were received by the destination than the ones sent by the source, then the route is not safe. For this purpose, a parameter named $P_{BH}(r)$ is defined. If the number of blocks in the acknowledgment packet for a route is fewer than the number of the blocks in the test packet or the acknowledgment packet is not received at all then the $P_{BH}(r)$ value will be increased for route r. However, if the test packet reaches the destination completely, then the received acknowledgment packet will contain the right number of blocks. This means that the route does not have a gray hole attack. In this case, the $P_{BH}(r)$ value will be decreased. In the proposed method, the test packet transmission process is carried out with a predefined number of blocks.

***The initial value of*** $P_{BH}(r)$ ***:*** If the route is verified according to the malicious node detection process, the initial value of $P_{BH}(r)$ will be set to zero. However, if the route is not verified, i.e. all of the data blocks in the test packet did not reach their destination, then $P_{BH}(r)$ will be set to 100. Afterward, based on whether or not it receives the acknowledgment packet, the source node updates the table for different routes according to Table 6. If the acknowledgment packet with the right number of blocks is received from the destination, $P_{BH}(r)$ will be decreased by 50. However, if the acknowledgment is not received from the destination or it contains the wrong number, then $P_{BH}(r)$ will be increased by 20. This process is repeated 2 times and $P_{BH}(r)$ is updated for all of the routes. Then if the $P_{BH}(r)$ value is higher than 50 for a route, that route will be identified as infected. The malicious nodes need to be discovered for the routes labeled as infected. This is carried out in the next phase of the proposed method. If the $P_{BH}(r)$ value is lower than 50, then the route is valid.

**Table 6** Route testing.

| RREP | Valid acknowledgment is received | Acknowledgment is not received or it is invalid | $P_{BH}(r)$ |
|---|---|---|---|
| $RREP_1$ | * | | $P_{BH}(r) - 50$ |
| $RREP_2$ | | * | $P_{BH}(r) + 20$ |

After testing the routes, the source IDS node starts discovering the malicious gray hole nodes along the infected routes.

**4.4 Phase 3: gray hole thing discovery in MTISS-IoT**
As mentioned before, in the gray hole attack, the gray hole nodes eliminate some of the packets they receive. A data control packet is proposed using this characteristic to check the intermediate nodes in an infected route. This packet includes three parameters as presented in Table 7.

**Table. 7** Fields of the Control Packet

| Node ID | ID – NEXT |
|---|---|
| Hash SHA – 256 | |

The fields of the proposed data control packet in figure 6 are as follows:

> *Node ID:* This field refers to the ID of the node that created the packet.
> *ID-NEXT:* This field refers to the next node on the route to the destination.
> *Hash SHA-256:* In order to discover the malicious gray hole node on the infected route, the source creates an encrypted message using the SHA-256 hashing function and puts the encrypted message in this field. This message must be identical in all of the data packets along the route. Each node needs to calculate the output once it receives the message and then forwards it to the source IDS node and also send it forward with its own specifications. A hashing function gets a string of digits and letters and generates a unique output with a fixed length. The proposed control packet is some kind of data packet. Therefore, the malicious nodes usually eliminate this packet, just like other packets, and do not send it to normal nodes. Also, these malicious nodes are unable to send the correct output to the source node.

Hashing functions have some important properties:

> 1. Their output does not change for a specific input string. If you give the same message to the hashing function over and over again, you will get the same output every time.
> 2. Finding the reverse transform is very difficult. If the output for a hashing function is available, it is almost impossible to find the corresponding input.
> 3. Finally, the slightest change in the input will change the output entirely. Therefore, even if a gray hole node does not eliminate the packet, it will be unable to generate the correct output and send it forward.

After discovering the infected routes, the source IDS node initiates the malicious gray hole node discovery process by generating a hash function. Then it sends the data control packet to the next hop along the route ($ID - NEXT$). Upon receiving the data control packet, each node needs to extract the hash function and generate a new data control packet with its own specifications. Then, send the hash function output back to the source node and send the new control packet to the next node. The output of the data packet which is sent along the reverse route is considered as the response for the original data packet. Once this response is received, if the received output is the same as the encrypted message then the source node updates its monitoring table and increases the $RREQ\_C$ and $RREQ\_T$ columns corresponding to that node by one. This is because the data packets have been sent correctly. This process is repeated along the route to the destination until one of the following occurs:

> 1. The data packet reaches the destination: In this case, the destination sends the ACK message with the correct output back to the source node along the reverse route. This means that the route is safe and there are no malicious nodes along this route.
> 2. The received output is not the same as the encrypted message that was sent: In this case, the next hop node which has sent the output message (ID-Next) will be identified as the malicious node. The source node will identify this malicious node by its ID.
> 3. The next intermediate node does not send a response for the data packet: This might be because either the node itself is malicious or the node before it is a malicious node and although it has sent a response back to the source, but it has prevented the message from being sent forward. Therefore, the source node needs to verify whether the node itself or the previous node is malicious. So, the source node asks both of the nodes to send their monitoring table and reviews their monitoring tables.

If the column $RREQ\_T$ is set to 1 in node $X$ and column $RREQ\_C$ is set to zero in node $Y$ (node $Y$ is the neighbor of node $X$) then node $Y$ is a malicious node. By carrying out this process, all of the malicious nodes on every route will be identified. These nodes must be introduced to the entire network.

**4.5 Phase 4: Malicious thing elimination process in MTISS-IoT**

In the proposed method, after the malicious nodes are discovered by the source IDS node, the source node broadcasts a message containing the ID of the malicious node in the network in order to remove that node from the routing process in the entire network. Once the malicious node is identified and separated from the network, all of the nodes separate any routing information containing the malicious node from their routes and future $RREP$ packets will not include the malicious node.

The proposed MTISS-IoT method tries to select the best and most reliable routes for data transmission by monitoring intermediate nodes. Furthermore, MTISS-IoT does not stop there and verifies the routes and intermediate nodes by sending test and control packets. The proposed method also discovers the malicious gray hole node and removes it from the routing process by using the monitoring table and sending encrypted control packets. The flowchart for the proposed MTISS-IoT method is presented in Figure 7.

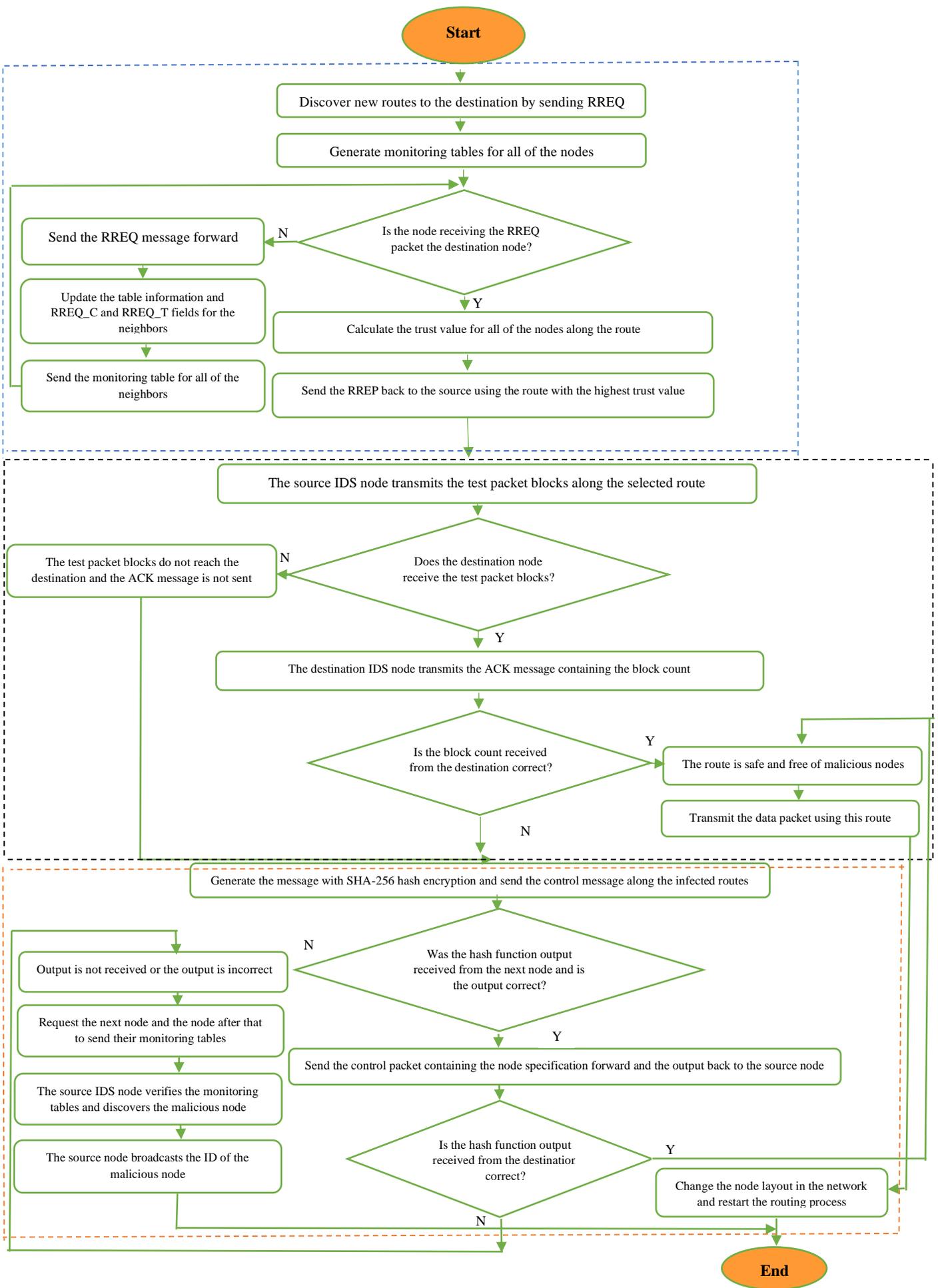

Fig. 7 Flowchart of the MTISS-IoT.

# 5 Performance evaluation

In following section, we show and discuss the experimental simulation results of the MTISS-IoT to prevent gray hole attack.

## 5.1 Performance metrics

In following section, we investigate the performance and effectiveness of the MTISS-IoT by a numerical simulation in NS-2. The results are compared with REATO and IRAD, and IMLADS approaches proposed in [13], [14], and [15] respectively. The notations of the considered for MTISS-IoT are as Table 8.

**Table 8** Abbreviated notations

| Parameters | Description |
|---|---|
| FPR | False positive rate |
| FNR | False negative rate |
| TPR | True positive rate |
| TNR | True negative rate |
| DR | Detection rate |

*FPR:* The FP is calculated by the total number of things wrongly detected as the malicious things divided by the total number of normal things [16-18]. Therefore, the *FPR* is defined as illustrated in Eq. (5).

$$FPR = \left(\frac{FPR}{FPR+TNR}\right)*100 \quad \text{Where:} \quad TNR = \left(\frac{TNR}{TNR+FPR}\right)*100 \tag{5}$$

*FNR:* The rate of the malicious thing to total normal things that were mistakenly marked as a normal thing [19-21]. Eq. (6) demonstrates the *FNR* calculation.

$$FNR = \left(\frac{TPR+TNR}{All}\right)*100 \quad \text{Where:} \quad TPR = \left(\frac{TPR}{TPR+FNR}\right)*100 \tag{6}$$

*DR:* It is determined as the ratio of the number of gray hole attack nodes marked to the total number of existing gray hole attack nodes in the IoT. *DR* is calculated by Eq. (7). Table 9 lists the parameters used for DR [22-27].

**Table 9** The parameters specified for *DR*

| Parameters | Description |
|---|---|
| True Positive (TP) | The *TP* is obtained from the whole number of marked gray hole attack nodes divided by the whole number of the gray hole attack nodes. |
| False Positive (FP) | The *FP* is obtained by the total number of nodes improperly recognized as the gray hole attack nodes divided by the whole number of normal nodes. |
| True Negative (TN) | The rate of the gray hole attack nodes being properly marked as a gray hole attack node. |
| False Negative (FN) | The rate of the gray hole attack node to whole normal sensors being wrongly marked as a normal node. |

$$DR = \left(\frac{TPR}{TPR + FNR}\right)*100 \quad \text{where} \quad All = TPR + TNR + FPR + FNR \tag{7}$$

## 5.2 Simulation setup and comparing algorithms

Because of the difficulty in debugging and implementing IoTs in real networks, it is necessary to view simulations as a basic design tool. The primary benefit of simulation is that analysis is simplified and protocol is verified, mostly, it is evident in systems in large scales [28-35]. The performance of the suggested method is assessed in this part by the use of NS-3 as the simulation means, and the discussion on the obtained results is presented. It should be noted that it is assumed that all REATO, MTISS-IoT, IMLADS, and IRAD settings and parameters are equal.

## 5.3 Simulation results and Analysis

In this section, we analyze the security performance of MTISS-IoT under the four attack scenarios (described in Table 10). These attack is categorized into DoS attack. There are 500 IoT things uniformly deployed in the network area initially. Some important parameters are listed in Table 10.

**Table 10** Setting of simulation parameters.

| Parameters | Value |
|---|---|
| Simulation tool | NS-3 |
| MAC | IEEE 802.11 |
| Transport | UDP/IPv6 |
| Wireless transmission range | 50 metres |
| Traffic type | CBR |
| Number of nodes, and Packet size | 500, 256 bytes |

The main simulation settings for four scenario are summarized in Table 11.

**Table 11** The setting of simulation parameters for four scenario.

| Scenario #1 | | Scenario #2 | |
|---|---|---|---|
| Malicious things rate | 8% | Malicious things rate | 16% |
| Coverage area (m x m) | 60 x 60 | Coverage area (m x m) | 70 x 70 |
| Simulation time | 500 | Simulation time | 1000 |
| **Scenario #3** | | **Scenario #4** | |
| Malicious things rate | 24% | Misbehaving things rate | 0, 0.05, 0.10, 0.15, 0.20, 0.25, 0.30 |
| Coverage area (m x m) | 80 x 80 | Coverage area (m x m) | 90 x 90 |
| Simulation time | 1500 | Simulation time | 2000 |

Table 12-14 compares the performance of MTISS-IoT with that of REATO, IRAD, and IMLADS in terms of $FPR$, $FNR$, and $DR$.

**Table 12** *FPR* vs misbehaving thing ratio.

| Misbehaving thing ratio | FPR (%) | | | |
|---|---|---|---|---|
| | IMLADS | IRAD | REATO | MTISS-IoT |
| 0 | 8.2 | 8.3 | 9.25 | 7.8 |
| 0.05 | 17.5 | 10.31 | 12.24 | 8.3 |
| 0.10 | 24.7 | 19.05 | 19.01 | 8.8 |
| 0.15 | 34.67 | 27 | 23.35 | 9.3 |
| 0.20 | 43.2 | 34.38 | 28.62 | 10.11 |
| 0.25 | 56.3 | 47.6 | 31.88 | 10.5 |
| 0.30 | 63.2 | 52.67 | 39.09 | 11.5 |

**Table 13** *FNR* vs misbehaving thing ratio.

| Misbehaving thing ratio | FNR (%) | | | |
|---|---|---|---|---|
| | IMLADS | IRAD | REATO | MTISS-IoT |
| 0 | 19.5 | 7.93 | 9.005 | 7.8 |
| 0.05 | 20.1 | 8.43 | 10.08 | 8.3 |
| 0.10 | 24.6 | 10.19 | 11.3 | 8.8 |
| 0.15 | 30.1 | 15.63 | 13.37 | 9.3 |
| 0.20 | 34.2 | 24.38 | 16.25 | 10.11 |
| 0.25 | 38.3 | 33.2 | 18.76 | 10.5 |
| 0.30 | 46.3 | 39.27 | 24.89 | 11.5 |

**Table 14** *DR* vs misbehaving thing ratio.

| Misbehaving thing ratio | DR (%) | | | |
|---|---|---|---|---|
| | IMLADS | IRAD | REATO | MTISS-IoT |
| 0 | 76.14 | 91.63 | 90.2 | 97.4 |
| 0.05 | 74.6 | 89.49 | 88.57 | 96.5 |
| 0.10 | 72.5 | 80.46 | 81.8 | 96.2 |
| 0.15 | 65.5 | 73.35 | 76.37 | 95.6 |
| 0.20 | 60.32 | 63.19 | 70.43 | 94.4 |
| 0.25 | 46.4 | 50.34 | 66.16 | 93.2 |
| 0.30 | 40.6 | 46.14 | 60.67 | 91.4 |

Average values of all methods for all metrics under gray hole attack are shown Table 15.

**Table 15** Average values of all methods (24% malicious things).

| Methods | FPR (%) | FNR (%) | DR (%) |
|---|---|---|---|
| MTISS-IoT | 17.443 | 20.37 | 82.06 |
| REATO | 26.392 | 27.23 | 65.91 |
| IRAD | 29.77 | 31.058 | 57.134 |
| IMLADS | 49.59 | 45.41 | 52.93 |

According to Table 16, the average delay of all approaches is large at the first; however, after a while and few iterations, the decision-making time is reduced. The reason for this reduction is that in this situation, the convergence and delay become static. Totally, based on this table, we can conclude that the delay of MTISS-IoT is the least among considered approaches.

**Table 16** Average delay (ms) comparison of MTISS-IoT with other approaches

| Time | MTISS-IoT | REATO | IRAD | IMLADS |
|---|---|---|---|---|
| T=100 | 11.5 | 39.4 | 42.4 | 46.3 |
| T=200 | 21.4 | 46.2 | 51.5 | 56.4 |
| T=300 | 20.2 | 41.68 | 48.67 | 51.2 |
| T=400 | 16.09 | 35.23 | 42.5 | 49.3 |

In the following, we evaluate the performance of the proposed MTISS-IoT method in the situation a gray hole attack occurs in the IoT network. The results have been presented in Table 17.

**Table 17** Average values for all metrics under gray hole attack

| Metrics (%) | Malicious node (%) | IMLADS (without attack) | IRAD (under attack) | REATO (under attack) | MTISS-IoT (under attack) |
|---|---|---|---|---|---|
| FPR | 30 | 60.2 | 49.67 | 36.09 | 14.4 |
| FPR | 40 | 65.8 | 53.19 | 40.49 | 16.3 |
| FNR | 30 | 43.3 | 36.27 | 21.89 | 9.3 |
| FNR | 40 | 49.1 | 42.72 | 29.28 | 12.2 |
| DR | 30 | 37.6 | 43.14 | 57.67 | 84.2 |
| DR | 40 | 32.1 | 38.68 | 51.39 | 81.94 |

**FPR:** Figure 8 presents the FPR of nodes participating in data sending/receiving operations in MTISS-IoT, REATO, IRAD, and IMLADS methods based on network topology. As observed in Fig. 8(a), the FPR produced by the suggested design showed small and mild growth in comparison with the other three designs if the malicious things' rate is raised from 8 to 24 percent and the range of normal things is from 50 to 500. The FPR of the suggested MTISS-IoT is below 12 percent while the number of normal things is 500, and the malicious things' rate is 8 percent. Nevertheless, it is set to 47 percent for IMLADS, 24 percent for the REATO, and it is 31 percent for the IRAD. The suggested design is superior because of the quick discovery of malicious things and their elimination by normal things and source thing cooperation. Moreover, it is also because the suggested algorithm discovers gray hole attack and separates them from the IoT network. Thus, the FPR, which happens due to attacks, is reduced. As shown in the figure, MTISS-IoT decreases the FPR by more than 17, 26, and 32% those of REATO, IRAD, and IMLADS models, respectively.

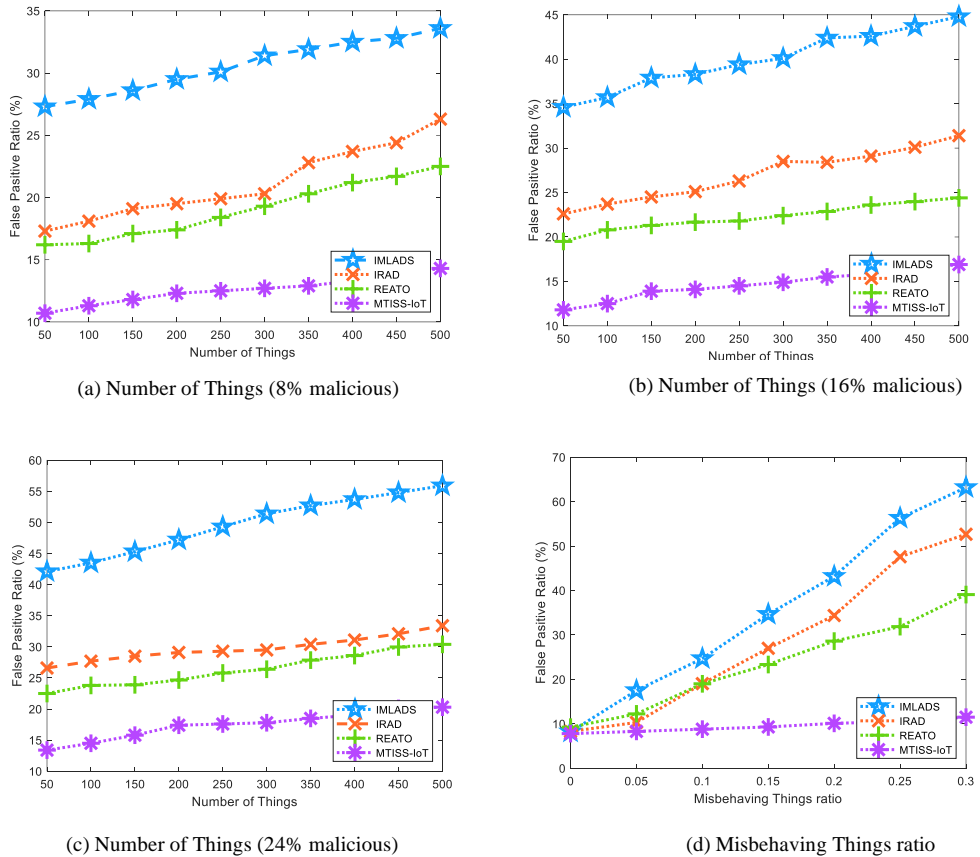

(a) Number of Things (8% malicious)

(b) Number of Things (16% malicious)

(c) Number of Things (24% malicious)

(d) Misbehaving Things ratio

**Fig. 8** Comparison of the MTISS-IoT proposed scheme, REATO, IRAD, and IMLADS approaches in term of FPR.

**FNR:** Figure 9 shows the comparison of the MTISS-IoT proposed scheme, REATO, IRAD, and IMLADS models in term of FNR in gray hole attack. (a) Number of Things (8% malicious), (b) Number of Things (16% malicious), (c) Number of Things (24% malicious), and (d) Misbehaving things ratio respectively. As indicated by the diagrams, the MTISS-IoT design's FNR has raised a little. However, its value is so higher in the IMLADS, IRAD, and REATO. As observed in Fig. 9(a), when the number of normal things is 500, the suggested design's FNR is below 18 percent. However, it is 25, 30, and 40 percent for the other three methods. In Fig. 9(b), when the malicious things' rate is 16 percent, it is below 22 percent in the suggested schema while it is 24, 27, and 46 percent for the other three approaches. In Figure 9(c), and (d) we observe that the adaptation capability of MTISS-IoT is higher than that of other approaches. This superior performance can be attributed to mainly, MTISS-IoT detection scheme.

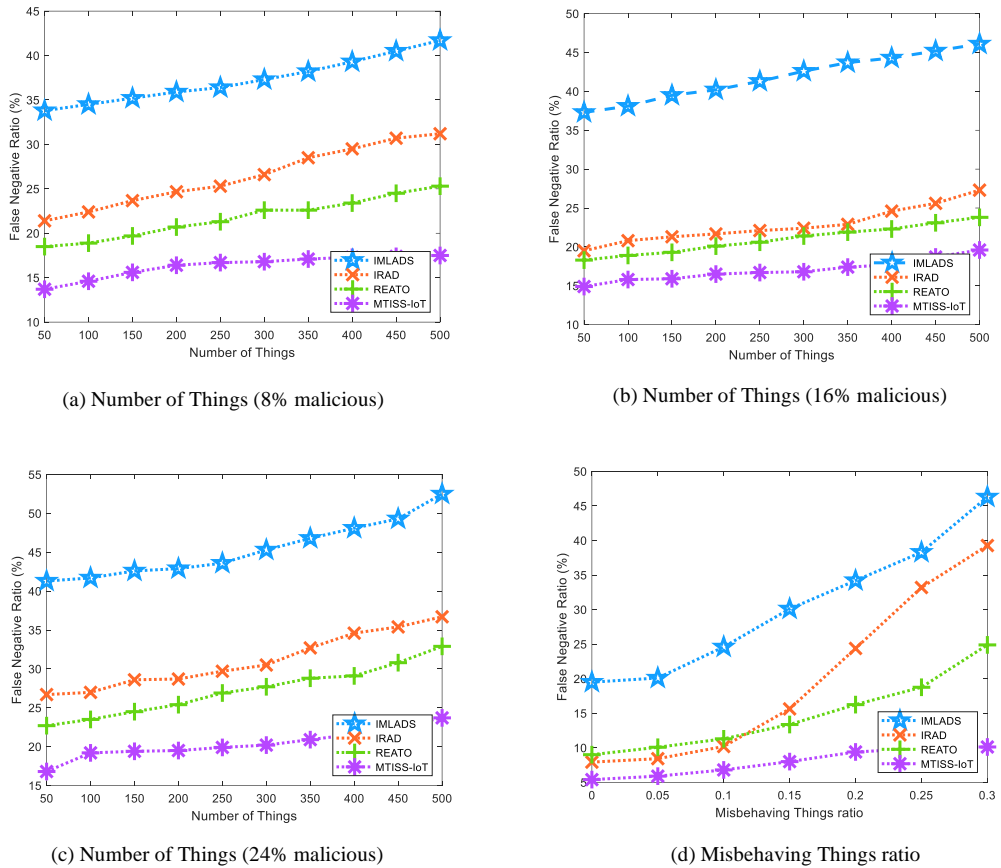

(a) Number of Things (8% malicious)

(b) Number of Things (16% malicious)

(c) Number of Things (24% malicious)

(d) Misbehaving Things ratio

**Fig. 9** Comparison of the MTISS-IoT proposed scheme, REATO, IRAD, and IMLADS approaches in term of FNR.

**DR:** Figure 10 shows the comparison of the MTISS-IoT proposed scheme, REATO, IRAD, and IMLADS models in term of DR. (a) Number of Things (8% malicious), (b) Number of Things (16% malicious), and (c) Number of Things (24% malicious), and (d) Misbehaving things ratio respectively. The diagrams indicate that the detection rate in the three approaches is decreased according to two scenarios, particularly when there are a high number of attacks. This decrease is evident more in the REATO compared to other mechanisms. The suggested schema is able to discover all of these attacks with a detection rate above 87.5 percent. When the number of normal things is 500 and the malicious things' rate is 16 percent, this result can be realized. When there are 24% malicious nodes in the IoT network which are using MTISS-IoT, the DR is about 83.13 % as shown in Fig. 10. In the case of REATO, IRAD, and IMLADS, the DR for all Number of Things is about 71%, 61%, and 56%, respectively.

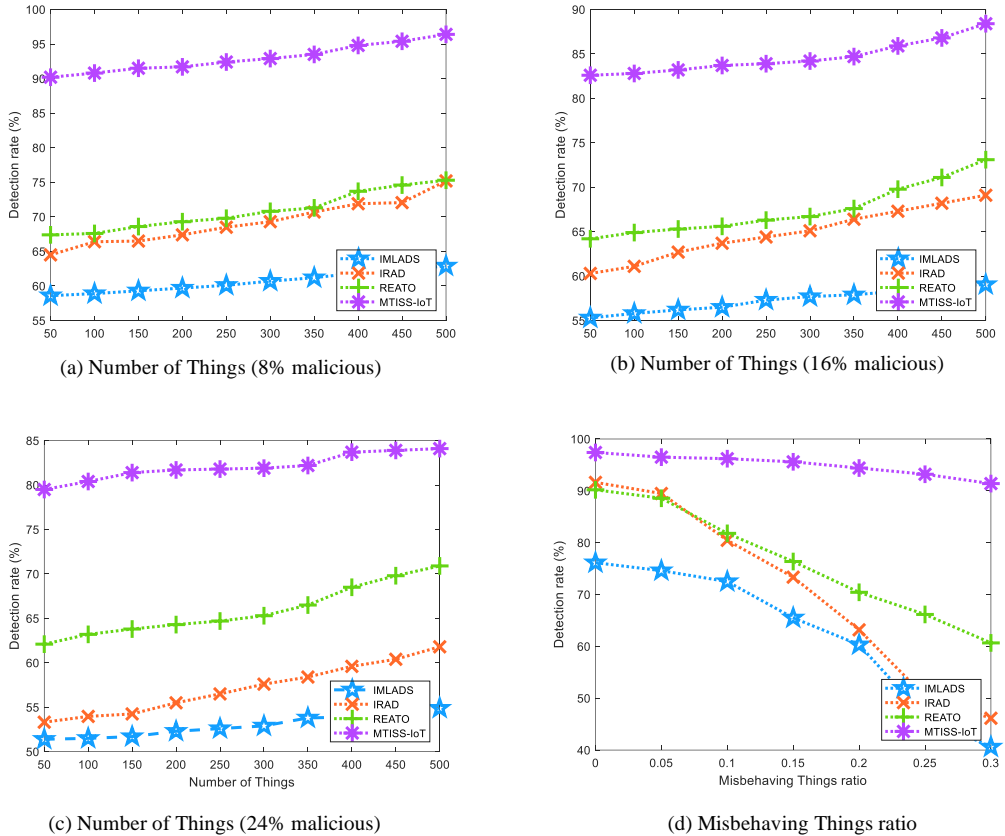

**Fig. 10** Comparison of the MTISS-IoT proposed scheme, REATO, IRAD, and IMLADS approaches in term of Detection rate.

The comparison results of the MTISS-IoT, in terms of PDR at different percent of gray hole attack are provided in Figure 11.

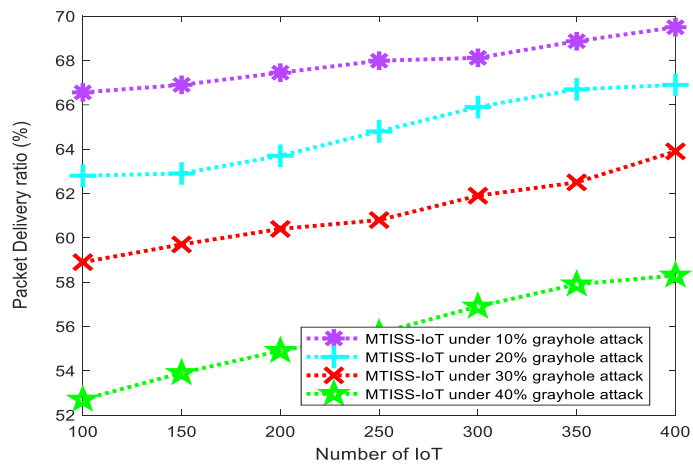

**Fig. 11** PDR vs different number of nodes at different percent of gray hole attack in MTISS-IoT

## 6 Conclusion and Future work

By an increase in the use of IoT as well as to an easy implementation of these networks, these networks are being increased by day-to-day. Therefore, the security was known as a necessary need for providing the protected communications among IoT nodes. In order to overcome the challenges, there is a need to create a secure novel intelligent agent-based strategy achieving both vast protected mode and the performance of the desired networks. In this study, a multi-level trust-based intelligence schema using the cryptographic authentication was proposed to avoid gray hole attacks in IoT. The proposed scheme first verifies the trust level for the nodes in the network and then discovers and eliminates the malicious gray hole nodes using control packets. We investigated the MTISS-IoT scheme performance using NS-3. According to the results of the simulation, the MTISS-IoT was highly powerful against gray hole attack. It was demonstrated that it enjoys a low FPR (below 11.104%) and a high level of security, high detection rate (above 94.50%), and low FNR (below 17.49%) in comparison with present methods. In future work, the use of Firefly optimization is suggested to further reduce consumption energy and malicious attacks on the internet of things. Firefly algorithm is proposed to cluster nodes and authenticate in two levels to prevent from attacks.

# Reference

1. Airehrour, D., Gutierrez, J., & Ray, S. K. (2016). Secure routing for internet of things: A survey. Journal of Network and Computer Applications, 66, 198-213.
2. Prabhakar, A., & Anjali, T. (2019). Gray Hole Attack as a Byzantine Attack in a Wireless Multi-Hop Network. Journal of Applied Security Research, 1-30.
3. Behzad, S., Fotohi, R., & Dadgar, F. (2015). Defense against the attacks of the black hole, gray hole and wormhole in MANETs based on RTT and PFT. International Journal of Computer Science and Network Solutions (IJCSNS), 3, 89-103.
4. Fotohi, R. (2020). Securing of Unmanned Aerial Systems (UAS) against security threats using human immune system. Reliability Engineering & System Safety, 193, 106675.
5. Bostani, H., & Sheikhan, M. (2017). Hybrid of anomaly-based and specification-based IDS for Internet of Things using unsupervised OPF based on MapReduce approach. Computer Communications, 98, 52-71.
6. Sahay, R., Geethakumari, G., Mitra, B., & Thejas, V. (2018, December). Exponential Smoothing based Approach for Detection of Blackhole Attacks in IoT. In 2018 IEEE International Conference on Advanced Networks and Telecommunications Systems (ANTS) (pp. 1-6). IEEE.
7. Hatzivasilis, G., Papaefstathiou, I., & Manifavas, C. (2017). SCOTRES: secure routing for IoT and CPS. IEEE Internet of Things Journal, 4(6), 2129-2141.
8. Airehrour, D., Gutierrez, J., & Ray, S. (2017). A trust-aware RPL routing protocol to detect blackhole and selective forwarding attacks.
9. Raoof, A., Matrawy, A., & Lung, C. H. (2018). Routing Attacks and Mitigation Methods for RPL-Based Internet of Things. IEEE Communications Surveys & Tutorials, 21(2), 1582-1606.
10. Bhalaji, N., Hariharasudan, K. S., & Aashika, K. (2019, January). A Trust Based Mechanism to Combat Blackhole Attack in RPL Protocol. In International Conference on Intelligent Computing and Communication Technologies (pp. 457-464). Springer, Singapore.
11. Patel, H. B., & Jinwala, D. C. (2019, October). Blackhole Detection in 6LoWPAN Based Internet of Things: An Anomaly Based Approach. In TENCON 2019-2019 IEEE Region 10 Conference (TENCON) (pp. 947-954). IEEE.
12. Conti, M., Kaliyar, P., Rabbani, M. M., & Ranise, S. (2018, October). SPLIT: A Secure and Scalable RPL routing protocol for Internet of Things. In 2018 14th International Conference on Wireless and Mobile Computing, Networking and Communications (WiMob) (pp. 1-8). IEEE.
13. Sicari, S., Rizzardi, A., Miorandi, D., & Coen-Porisini, A. (2018). REATO: REActing TO Denial of Service attacks in the Internet of Things. Computer Networks, 137, 37-48.
14. Yavuz, F. Y., Ünal, D., & Gül, E. (2018). Deep learning for detection of routing attacks in the internet of things. International Journal of Computational Intelligence Systems, 12(1), 39-58.
15. Qin, T., Wang, B., Chen, R., Qin, Z., & Wang, L. (2019). IMLADS: Intelligent maintenance and lightweight anomaly detection system for internet of things. Sensors, 19(4), 958.

16. Jamali, S., & Fotohi, R. (2017). DAWA: Defending against wormhole attack in MANETs by using fuzzy logic and artificial immune system. the Journal of Supercomputing, 73(12), 5173-5196.
17. Fotohi, R., Bari, S. F., & Yusefi, M. (2019). Securing Wireless Sensor Networks Against Denial-of-Sleep Attacks Using RSA Cryptography Algorithm and Interlock Protocol. International Journal of Communication Systems.
18. Sarkohaki, F., Fotohi, R., & Ashrafian, V. (2017). An efficient routing protocol in mobile ad-hoc networks by using artificial immune system. International Journal of Advanced Computer Science and Applications (IJACSA), 8 (4).
19. Alcaraz, C. (2019). Security and Privacy Trends in the Industrial Internet of Things. Springer.
20. Fotohi, R., Ebazadeh, Y., & Geshlag, M. S. (2016). A new approach for improvement security against DoS attacks in vehicular ad-hoc network. International Journal of Advanced Computer Science and Applications, 7(7), 10-16.
21. Rathee, G., Sharma, A., Kumar, R., & Iqbal, R. (2019). A Secure Communicating Things Network Framework for Industrial IoT using Blockchain Technology. Ad Hoc Networks, 101933.
22. Behzad, S., Fotohi, R., Balov, J. H., & Rabipour, M. J. (2018). An Artificial Immune Based Approach for Detection and Isolation Misbehavior Attacks in Wireless Networks. JCP, 13(6), 705-720.
23. Zhang, J., Rajendran, S., Sun, Z., Woods, R., & Hanzo, L. (2019). Physical Layer Security for the Internet of Things: Authentication and Key Generation. IEEE Wireless Communications.
24. Fotohi, R., & Jamali, S. (2014). A comprehensive study on defence against wormhole attack methods in mobile Ad hoc networks. International journal of Computer Science & Network Solutions, 2, 37-56.
25. Butun, I., Österberg, P., & Song, H. (2019). Security of the Internet of Things: Vulnerabilities, Attacks and Countermeasures. IEEE Communications Surveys & Tutorials.
26. Zandiyan S, Fotohi R, Koravand M. P-method: Improving AODV routing protocol for against network layer attacks in mobile Ad-Hoc networks. International Journal of Computer Science and Information Security. 2016 Jun 1;14(6):95.
27. Fotohi, R., Heydari, R., & Jamali, S. (2016). A Hybrid routing method for mobile ad-hoc networks. Journal of Advances in Computer Research, 7(3), 93-103.
28. Xia, J., Xu, Y., Deng, D., Zhou, Q., & Fan, L. (2019). Intelligent secure communication for Internet of Things with statistical channel state information of attacker. IEEE Access, 7, 144481-144488.
29. Fotohi, R., Jamali, S., Sarkohaki, F., & Behzad, S. (2013). An Improvement over AODV routing protocol by limiting visited hop count. International Journal of Information Technology and Computer Science (IJITCS), 5(9), 87-93.
30. Jamali, S., & Fotohi, R. (2016). Defending against wormhole attack in MANET using an artificial immune system. New Review of Information Networking, 21(2), 79-100.
31. Zhang, T., Zhang, T., Ji, X., & Xu, W. (2019, July). Cuckoo-RPL: Cuckoo Filter based RPL for Defending AMI Network from Blackhole Attacks. In 2019 Chinese Control Conference (CCC) (pp. 8920-8925). IEEE.
32. Behzad, S., Fotohi, R., & Jamali, S. (2013). Improvement over the OLSR routing protocol in mobile Ad Hoc networks by eliminating the unnecessary loops. International Journal of Information Technology and Computer Science (IJITCS), 5(6), 2013.
33. Kandhoul, N., Dhurandher, S. K., & Woungang, I. (2019). T_CAFE: A Trust based Security approach for Opportunistic IoT. IET Communications.
34. Jamali, S., Fotohi, R., Analoui, M. (2018). An Artificial Immune System based Method for Defense against Wormhole Attack in Mobile Adhoc Networks. TABRIZ JOURNAL OF ELECTRICAL ENGINEERING, 47(4), 1407-1419.
35. Sani, A. S., Yuan, D., Jin, J., Gao, L., Yu, S., & Dong, Z. Y. (2019). Cyber security framework for Internet of Things-based Energy Internet. Future Generation Computer Systems, 93, 849-859.